\documentclass[final,3p,times,twocolumn]{elsarticle} 

\usepackage{lineno,amsthm,amsmath,siunitx,float,subfig,subfloat,textgreek,textcomp,tabularx}
\usepackage[T1]{fontenc}
\usepackage[bookmarksopen=true,colorlinks=true,urlcolor=blue]{hyperref}

\journal{NIM A}

\newcolumntype{l}{>{\hsize=0.2\hsize}X}
\newcolumntype{r}{>{\hsize=1.8\hsize}X}

\newcommand{\ignore}[1]{}

\begin{document}
	\begin{frontmatter}
		\title{The Silicon Sensors for the High Granularity Calorimeter of CMS}
		\author{Peter Paulitsch\corref{cor1}}
		\cortext[cor1]{Corresponding author}
		\ead{peter.paulitsch@cern.ch}
		\author{on behalf of the CMS Collaboration}
		\address{Austrian Academy of Sciences, Institute of High Energy Physics (HEPHY), Nikolsdorfer Gasse 18, 1050 Wien, Austria}
		
		\begin{abstract}
			The installation of the High-Luminosity Large Hadron Collider (\mbox{HL-LHC}) presents unprecedented challenges to experiments like the Compact Muon Solenoid (\mbox{CMS}) in terms of event rate, integrated luminosity and therefore radiation exposures. To cope with this new environment, new detectors will be installed during the \mbox{CMS} Phase 2 Upgrade, including the replacement of the calorimeter endcaps with the "High Granularity Calorimeter" (\mbox{HGCAL}), which contains silicon sensors and scintillators as active elements. The silicon sensors will be produced in an 8" wafer process, which is new for high-energy physics, so it demands extensive quality verification. \ignore{The silicon sensors contain hexagonally shaped n-in-p DC-coupled diodes, which allow for maximization of the usable wafer area. }A first batch of prototype sensors underwent electrical tests at the institutes of the \mbox{CMS} Collaboration. Testing revealed major problems with the mechanical stability of the thin backside protective layer, that were not seen in earlier 6" prototypes produced by a different backside processing method. Following these results, the \mbox{HGCAL} group introduced the concept of "frontside biasing", allowing testing of the sensors without exposing its backside, verified the applicability, and adapted the prototype design to apply this method in series production.
		\end{abstract}
	
	\end{frontmatter}
	
	\section{Introduction}
\begin{figure}[htb]\centering
		\includegraphics[width=0.5\textwidth]{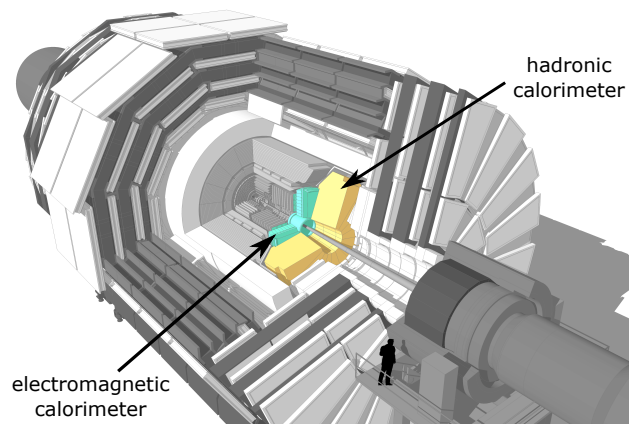}
		\caption{Location of the \mbox{HGCAL} at the \mbox{CMS} endcaps~\cite{CMSsubdetectorImages}.}
		\label{fig:location}
	\end{figure}

\begin{figure}[htb]\centering
		\includegraphics[width=0.5\textwidth]{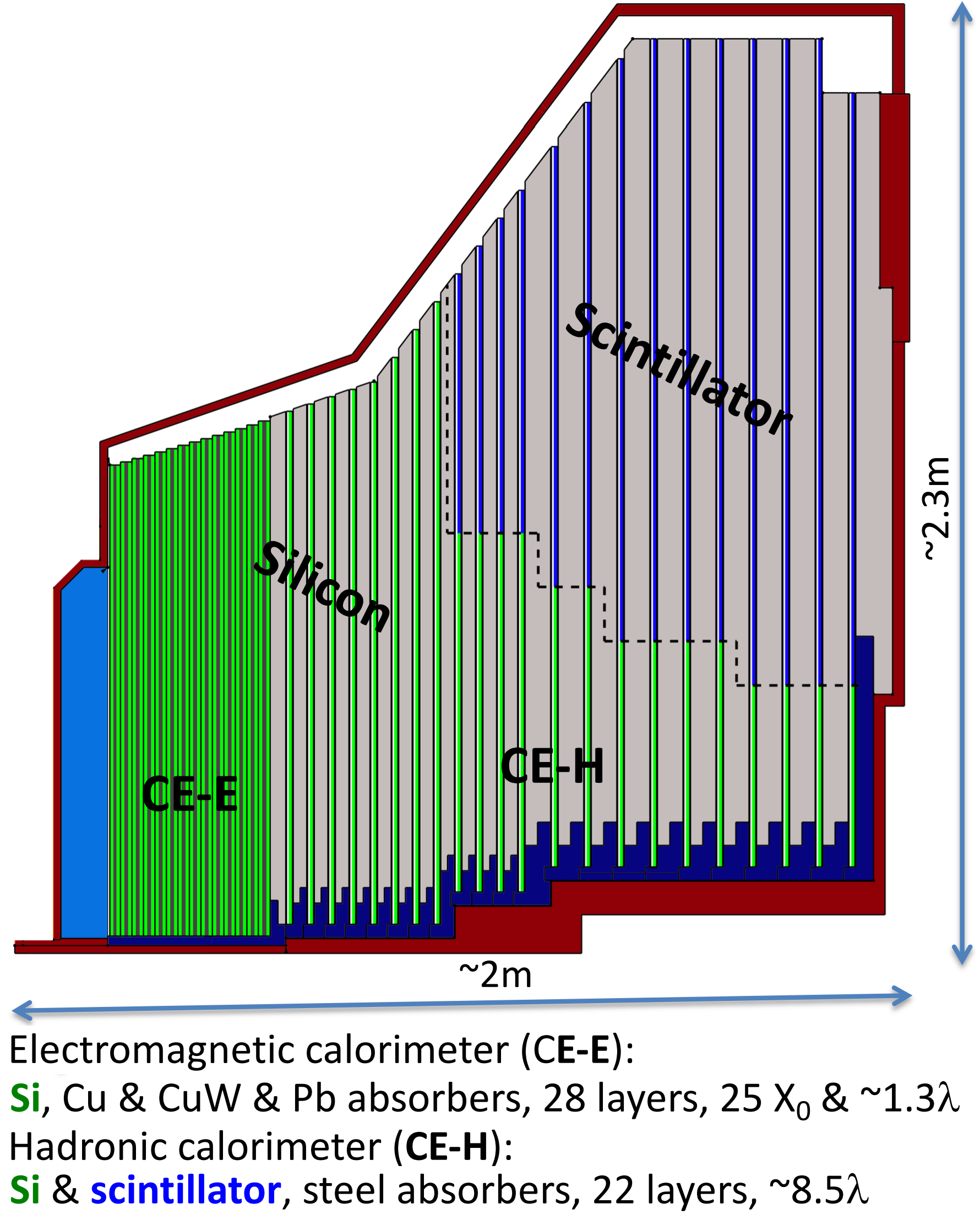}
		\caption{Schematic cross section of an endcap sector~\cite{HGCALoverview}.}
		\label{fig:schematic}
	\end{figure}

	During the Phase-2 Upgrade (2025 to 2027), the Large Hadron Collider (\mbox{LHC}) will be upgraded to the High-Luminosity \mbox{LHC} \mbox{(HL-LHC)} \cite{TDRHighLumiLHC}. The \mbox{HL-LHC} will have a factor 5--7 higher instantaneous luminosity compared to the end of \mbox{LHC} operation, resulting in a proportionally higher pileup and a factor 10 increase in integrated luminosity (\SI{3000}{\per\femto\barn}) over 10 years of operation. As a result, unprecedented levels of radiation and particle shower densities will affect experiments such as the Compact Muon Solenoid (\mbox{CMS}) \cite{CMSexperiment}.\ignore{The current sensors are already suffering from high radiation damage.} At these high collision rates, the overlap of particle showers will not be negligible any more, so detectors with increased spatial resolution are needed to distinct different showers. To address these challenges, the \mbox{CMS} Collaboration will upgrade its subdetectors including a replacement for the existing endcap calorimeters with the new High Granularity Calorimeter (\mbox{HGCAL}) \cite{TDRphase2HGCAL}, as shown in  Figures~\ref{fig:location} and~\ref{fig:schematic}. The calorimeter will utilize about 30000\footnote{This value is preliminary, it may be changed for optimized calorimeter coverage by partial sensors, see Chapter~\ref{subsec:FSB}} sensor modules covering more than \SI{620}{\square\metre}, allowing for efficient mitigation of pileup and particle-flow calorimetry.
	
	The \mbox{HGCAL} will be a sandwich calorimeter and will include an electromagnetic part (Calorimeter Endcap - Electromagnetic, \mbox{CE-E}) and a hadronic part (Calorimeter Endcap - Hadronic, \mbox{CE-H}). While the active sensing elements of the electromagnetic part will be entirely made of silicon sensors, the hadronic elements will implement silicon just for the inner high radiation domain. At the outer regions with lower radiation levels, plastic scintillators coupled to silicon photomultipliers~\cite{TDRphase2HGCAL} will be used, as shown in Figure~\ref{fig:schematic}. This article focuses on the silicon parts of the \mbox{CE-E} and \mbox{CE-H} sections.\\
	
	\section{Silicon sensors for the HGCAL}\label{sec:siliconHGCAL}
\ignore{\cite{TDRphase2tracker}}The silicon sensors of the \mbox{HGCAL} will be produced in an 8" process~\cite{TDRphase2HGCAL}, in contrast to earlier applications in high-energy physics, which used 6" processes. The leap towards the 8" process decisively reduces production costs and sensor testing efforts. However, this process is new to large-area sensors for high-energy physics and therefore brings new challenges in terms of radiation hardness, high-voltage stability, and other issues like sensor backside sensitivity (see Chapter~\ref{sec:sensorBackside}).
Three different sensor thicknesses will be deployed: \SI{120}{\micro\metre}, \SIlist{200;300}{\micro\metre}. Radiation damage increases the sensor leakage current. Thinner sensors draw less leakage current, so these will be utilized in regions with increased radiation levels to cope with higher radiation damage induced leakage currents (Table~\ref{tab:sensors}).

{\renewcommand{\arraystretch}{1.2}
	\begin{table}[ht]\centering
		\caption{Active thicknesses $d\textsubscript{act}$, number of channels, maximum expected fluences $\Phi\textsubscript{neq}$ (normalized to \SI{1}{\mega\electronvolt} neutron equivalent), and maximum expected total ionizing dose (TID) at \SI{3000}{\per\femto\barn}~\cite{TDRphase2HGCAL}. $\Phi\textsubscript{neq}$ is defined as the number of neutrons passing per sensor area and typically describes lattice displacement damage. The energy spectrum is normalized to \SI{1}{\mega\electronvolt} monoenergetic neutrons. The TID is the  dose by charged particles, and typically describes damages in oxide layers, caused by generating and accumulating immobile charges.}
		\begin{tabularx}{0.45\textwidth}{ccccc}
			\hline
			& full-size & & \\
			$d\textsubscript{act}$ (\si{\micro\meter}) & channels & $\Phi\textsubscript{neq}$ (cm\textsuperscript{-2}) & TID (\si{\gray})\\
			\hline
			120 & 432 (HD) & \num{7.0d15} & \num{1d6}\\
			200 & 192 (LD) & \num{2.5d15} & \num{2d5}\\
			300 & 192 (LD) & \num{5.0d14} & \num{3d4}\\
			\hline
		\end{tabularx}
		\label{tab:sensors}
\end{table}}

\begin{figure}[H]\centering
	\includegraphics[width=0.5\textwidth]{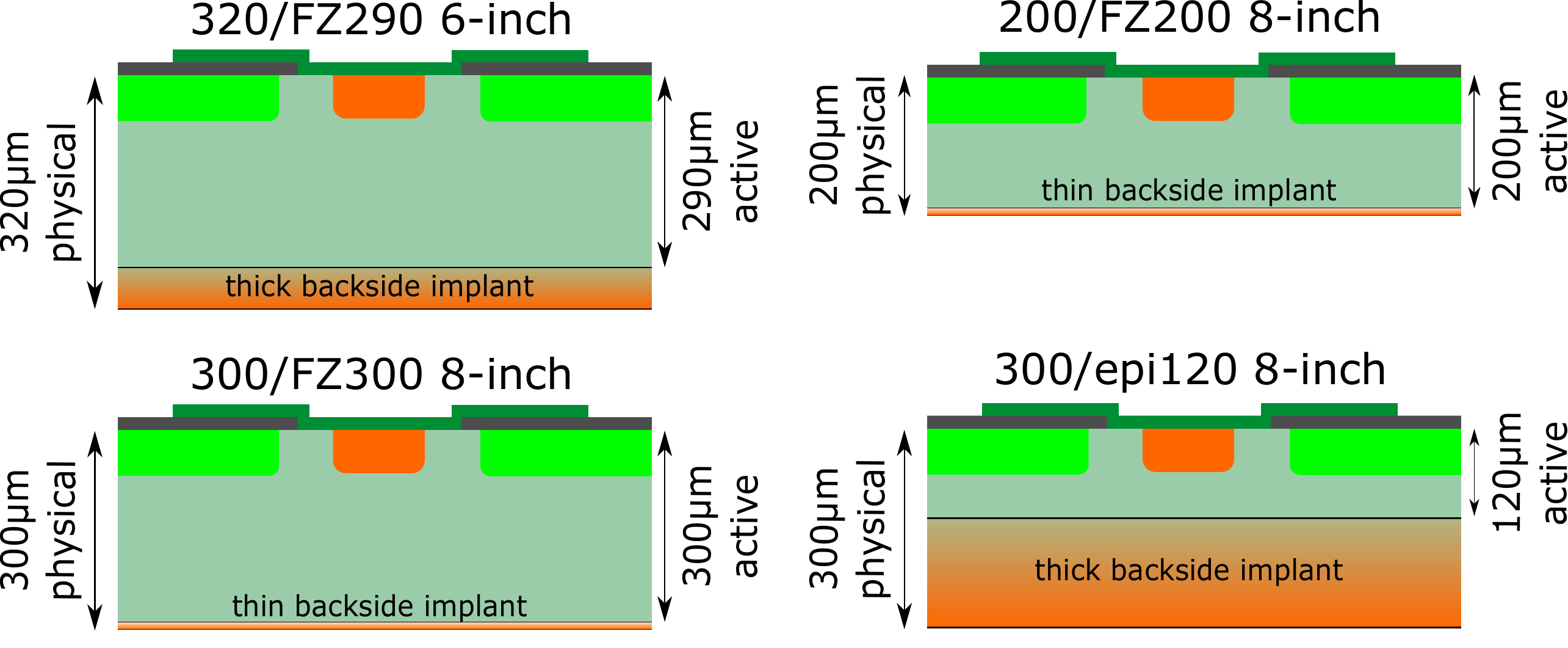}
	\caption{Sensor cross sections at different manufacturing processes. "epi" denotes an epitaxial process, "FZ" stands for "float-zone". For comparison, one 6" processed sensor cross section is shown. The thin backside implant of the FZ200 and FZ300 processes is about \SI{1}{\micro\metre} thick.}
	\label{fig:CS}
\end{figure}

\begin{figure}[htb]\centering
	\includegraphics[width=0.5\textwidth]{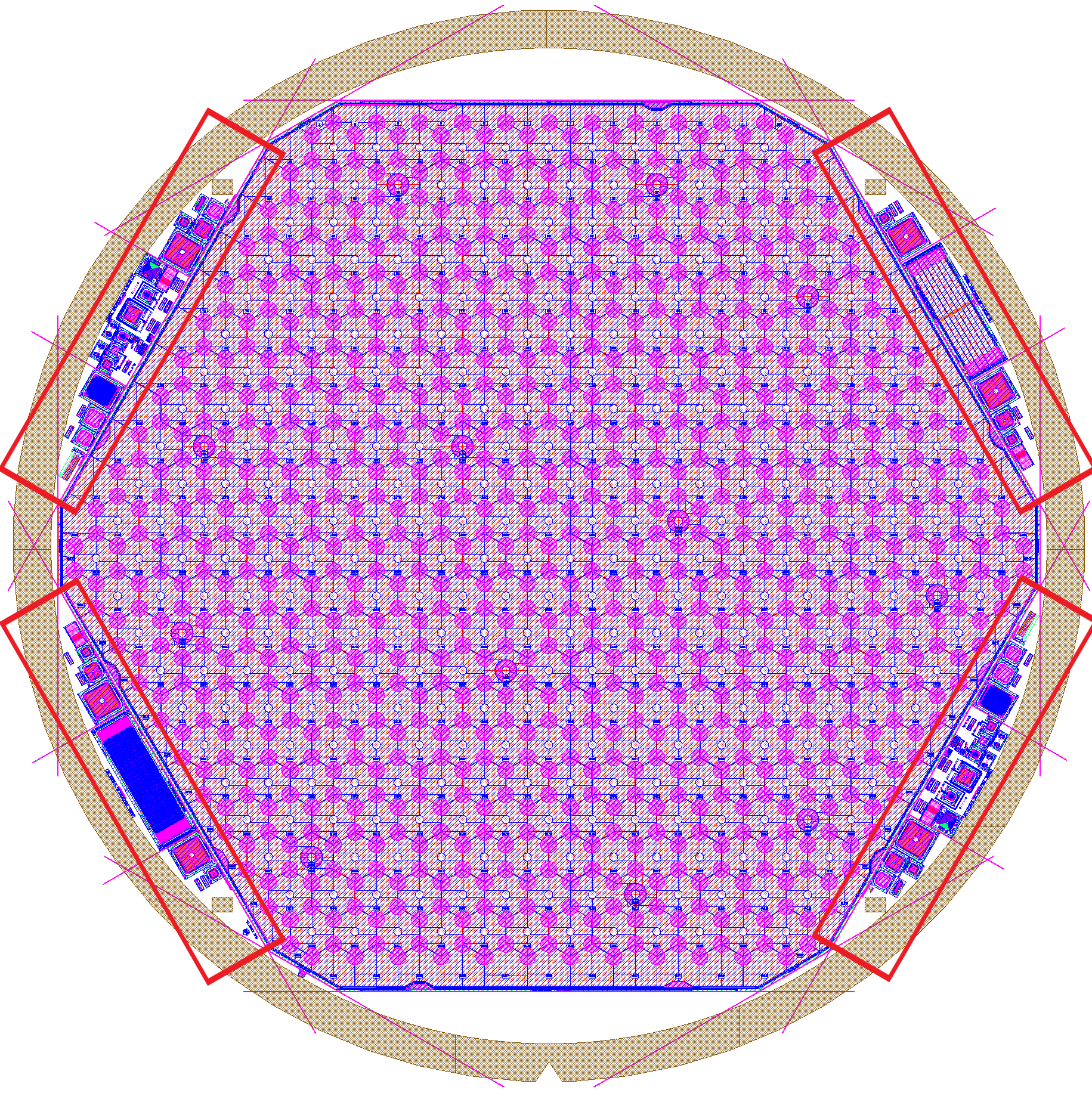}
	\caption{432-channel (HD) full wafer design, thickness \SI{120}{\micro\meter}. The red squares at the wafer periphery mark the four test structure half moons \cite{Hinger20PQC}.}
	\label{fig:HDfullContact}
\end{figure}

Hamamatsu will manufacture the \SI{120}{\micro\metre} sensors in an epitaxial process, whereas the thicker will be produced in a float-zone process, as shown in Figure~\ref{fig:CS}.\\The shape of a full sensor is hexagonal (see Figure~\ref{fig:HDfullContact}) because a hexagon is the largest seamlessly tileable, regular shape on a circular wafer. This maximizes the wafer-area usage, reduces the number of necessary sensor tiles, hence decreases costs.
Consequently, the active sensing elements also have a hexagonal shape, except for some irregular cells at the sensor edges and corners. Studies from \mbox{RD50}\footnote{Radiation hard semiconductor devices for very high luminosity colliders, \url{http://rd50.web.cern.ch/rd50/}} \cite{RD50chargeCollection} have shown that n-in-p materials have better charge collection than p-in-n, so like in the \mbox{CMS} Tracker \cite{TrackerPType}, n-in-p diodes were chosen also for the \mbox{HGCAL}. For the \SIlist{200;300}{\micro\metre} thick sensors, the so-called "Low-Density" (LD) design with 192 full-size channels is chosen. To keep the diode (cell) capacitances and therefor series noise in the preamplifier low, the \SI{120}{\micro\metre} thick sensors have smaller cells and therefore 432 full-size channels, the "High-Density" (HD) design as shown in Figure~\ref{fig:HDfullContact}.
To control the electric fields and to reduce the leakage current at the edge region, the sensor features two $n\textsuperscript{++}$ doped guard rings; the inner one is connected to ground and the outer one is floating. A $p\textsuperscript{++}$ doped edge ring is placed at the sensor edge, as shown in Figure~\ref{fig:extensions}. This edge ring brings the edge region to the same potential as the $p\textsuperscript{++}$ doped backside implant. Otherwise, microscopic cracks at the cutting edges would lead to field spikes which may cause breakdowns.

\section{Sensor testing}
\subsection{Test system}\label{subsec:probestation}
During the production phase, multiple \mbox{CMS} institutes will do quality control of 1--\SI{2}{\percent} of all produced sensors. Because of the large quantities of sensors produced, \mbox{CMS} needs fast and widely automatized test equipment. Probe-cards are a common standard in the semiconductor industry for large-scale quality control because they allow fast and easily reproducible measurements. Therefore the collaboration developed an open-source full wafer probe-card and switching system, called "\mbox{ARRAY}"~\cite{ARRAY}. A probe-card consists of a carrier which holds a set of probe tips in a defined geometric layout.\\Characterization of the sensors is primarily fulfilled by measuring current versus voltage (IV) and capacitance versus voltage (CV) curves \cite{BrondolinHGCALelectrical}\cite{AlmeidaVersatileSystems}. For all individual sensor cells on the same wafer, these measurements allow extracting parameters like the full depletion voltage ($V_{\text{fd}}$), full depletion capacitance ($C_{\text{fd}}$), or the breakdown voltage ($V_{\text{bd}}$).

\subsection{Sensor backside sensitivity and scratch tests}\label{sec:sensorBackside}
\begin{figure}[H]\centering
	\includegraphics[width=0.5\textwidth]{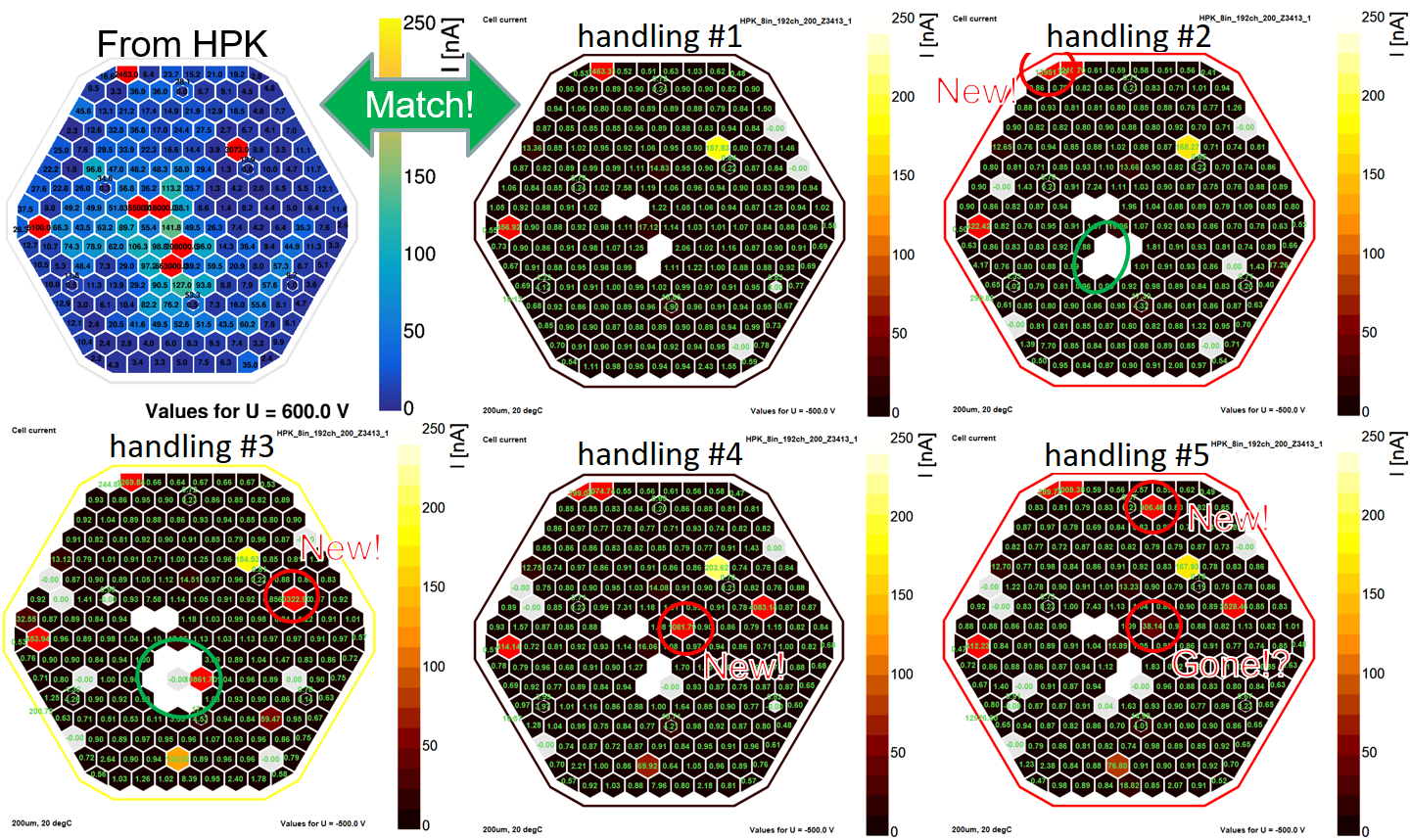}
	\caption{Sensor degradation after repeated handling steps, 192-channels, \SI{200}{\micro\metre} active thickness. New breakdown cells (marked by circles) appeared after each handling step. Cells with values of "0.00" have bad contacts. The first test ("From HPK") was done by Hamamatsu before shipment.}
	\label{fig:ManniWithHandling}
\end{figure}

\begin{figure}[htb]\centering
	\includegraphics[width=0.5\textwidth]{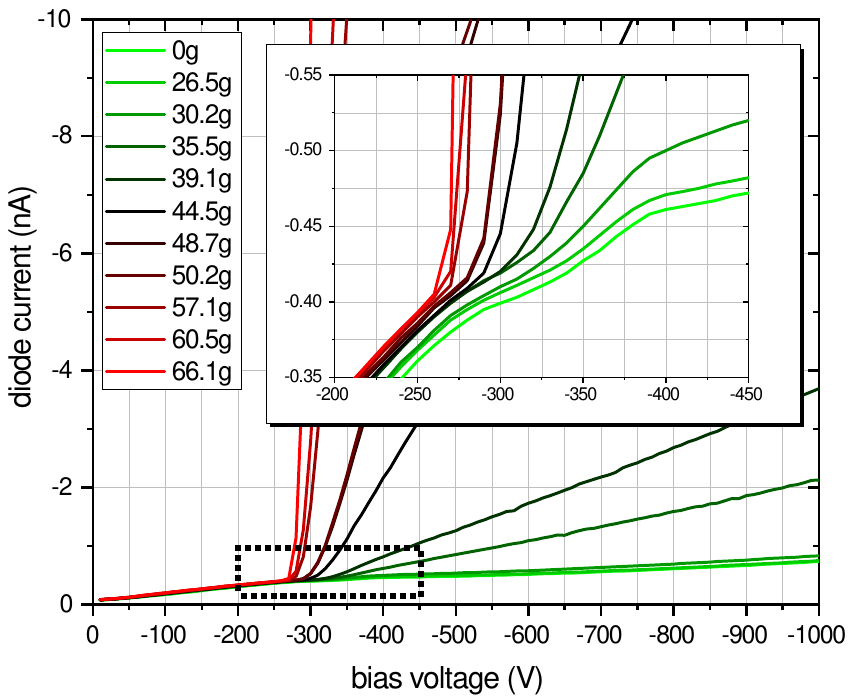}
	\caption{Test structure diode currents as a function of the bias voltage for increasing scratch depth, indicated by needle weight. The zoomed inlay shows a reduction of the breakdown voltage and increased currents with higher needle weight.}
	\label{fig:BrezIVscratches}
\end{figure}

\begin{figure}[htb]\centering
	\includegraphics[width=0.5\textwidth]{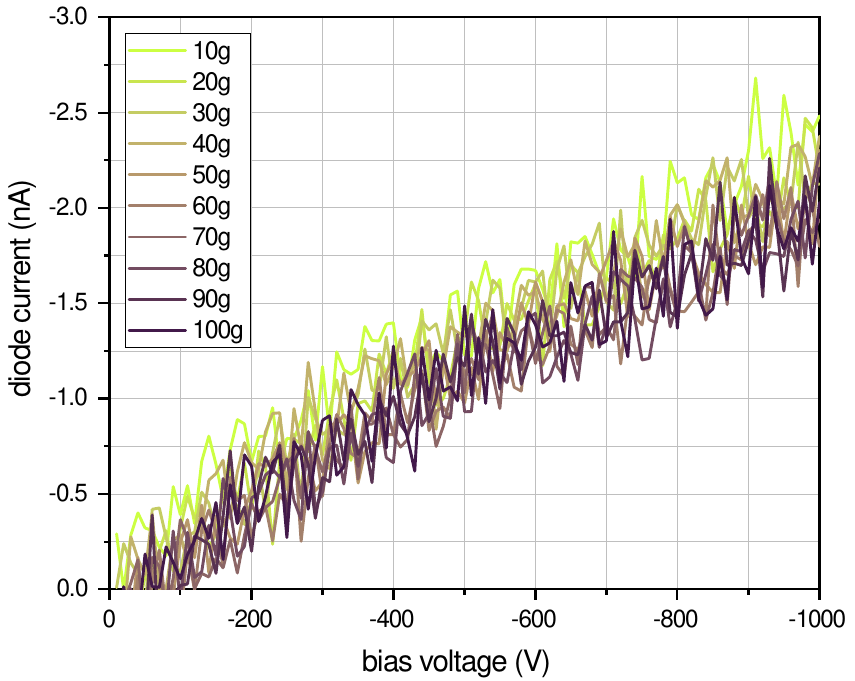}
	\caption{Test structure diode currents as a function of the bias voltage for increasing pin pressure, indicated by pin weight. No systematic behavior is seen at different pin pressures.}
	\label{fig:BrezPogoPin}
\end{figure}

Diode current characteristics on 8" \mbox{HGCAL} sensor prototypes showed a degradation in terms of an increasing number of cells with early breakdowns as shown in Figure~\ref{fig:ManniWithHandling}. This degradation only occured after sensor handling procedures were done. Repeated measurements without handling procedures in between remained stable. Because earlier 6" prototypes did not show this behavior, we suspected that the thin backside metalization of the 8" prototypes might cause early breakdowns and increased sensor currents. The previous 6" prototypes had a backside with a thick ("deep-diffused") field stop implant, seen on Figure~\ref{fig:CS}. The purpose of the highly doped field stop is to have zero electrical field at the beginning of the backside metalization. This stops the spread of the field also at high bias voltage.
To investigate these problems with the fragile backside, scratch tests with a tungsten carbide needle were performed on test structures~\cite{Hinger19PQC}\cite{PreeDissTS}. Using a needle manipulator, the needle tip (\SI{50}{\micro\metre} diameter) was kept hovering over the test structure diode, which was fixed to a precision scale (Kern 572-30, reproducibility \SI{1}{\micro\gram}). The underlying table was raised until the needle touched the sensor surface and the scale displayed the desired weight. Subsequently, the table was moved to create a scratch. After each scratch, the needle was removed and the diode IV characteristics were recorded, as shown in Figure~\ref{fig:BrezIVscratches}. These tests showed that the diode's backside is highly sensitive to scratches, the breakdown voltage decreased with scratches made at higher needle weights and diode currents increased at deeper scratches.
Depth measurements via laser interferometry (Wyko NT 3300 Profiling System) of these scratches showed that indeed the scratch depth increases with higher needle weight, and that all tried weights (\SI{10}{\gram} and more) penetrated the \SI{1.1}{\micro\metre} backside aluminum layer.\\Vertical pressure tests with variable pressures of the probe-card pins showed that the applied pressure of these pins did not alter the electrical characteristics of the diodes, even when measured during applied pressure, as shown in Figure~\ref{fig:BrezPogoPin}.

\subsection{Frontside biasing and edge/guard ring extensions}\label{subsec:FSB}
\begin{figure}[h]\centering
	\includegraphics[width=0.5\textwidth]{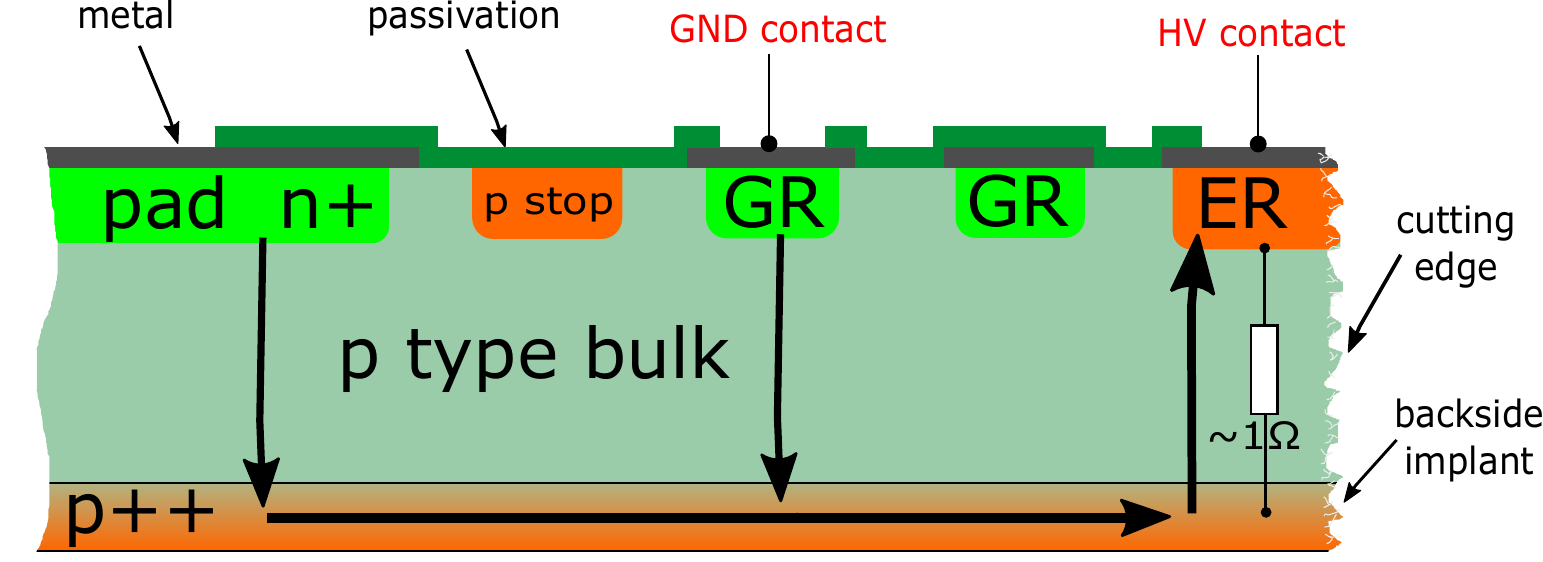}
	\caption{Sketch of cross section of the sensor edge using the frontside biasing concept. Current flow is visualized by the black arrows. Since there is a low resistance (\textasciitilde\SI{1}{\ohm}) path between edge ring and backside implant, the current flows via the backside to the edge ring, where the HV contact is applied.}
	\label{fig:FSBsketch}
\end{figure}

Initially, it was planned to apply the bias voltage during testing via the sensor's backside aluminum metalization. Due to the aforementioned backside fragility, Hamamatsu discussed with the collaboration to glue a compound polyimide (Kapton\texttrademark) foil on the backside during production to prevent scratches. This procedure would make it impossible to directly contact the backside. By exploiting the low-resistance path between edge ring, bulk and backside implant (all are p-doped, shown in Figure~\ref{fig:FSBsketch}), it is possible to use the edge ring as a contact for sensor biasing instead. This method is referred as "frontside biasing"~\cite{FSB2018}.

\begin{figure}[h]\centering
	\includegraphics[width=0.5\textwidth]{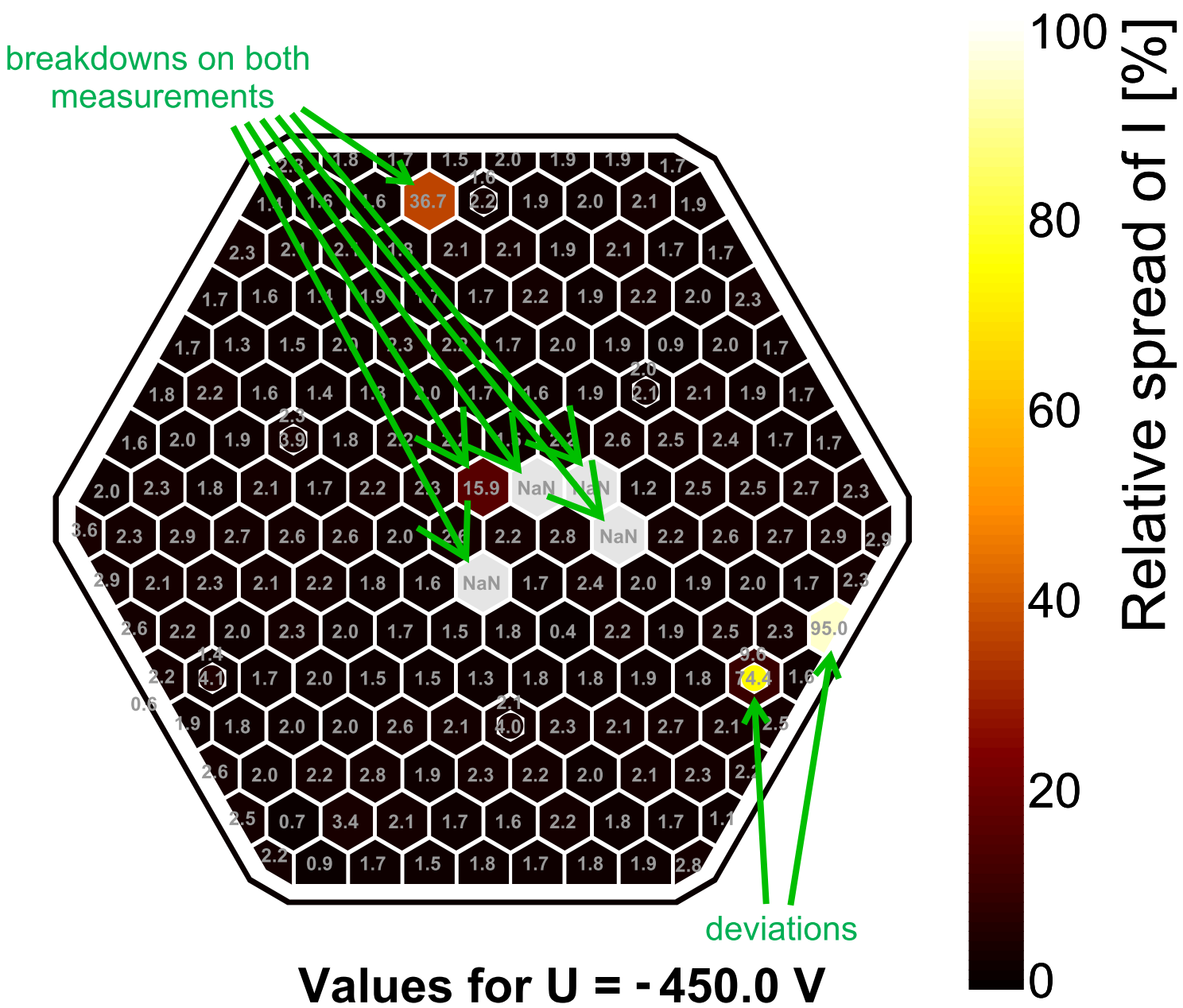}
	\caption{Cell current deviations between backside and frontside biasing of a \SI{200}{\micro\metre} thick sensor. Values are given in \% (Equation~\ref{eq:ratio}) at \SI{-450}{\volt}.}
	\label{fig:BsbFsb450V_unc_markers}
\end{figure}

To verify this on the 8" prototypes, we tested frontside biasing on an unirradiated, 192-channel \SI{200}{\micro\metre} thick sensor and compared frontside biasing cell currents to backside biasing cell currents. Figure~\ref{fig:BsbFsb450V_unc_markers} shows the relative spread ($R$) as derived in Equation~\ref{eq:ratio}, between front side bias current $I_{\text{FSB}}$ and back side bias current $I_{\text{BSB}}$ for each cell.

\begin{equation}
	R(\SI{}{\percent}):=\left|\frac{I_{\text{BSB}}}{a}-1\right|*\SI{100}{\percent}\quad \textrm{and} \quad a=\frac{I_{\text{BSB}}+I_{\text{FSB}}}{2}
	\label{eq:ratio}.
\end{equation}
For most cells not in breakdown, the differences are below \SI{2}{\percent}. There were two exceptions with remarkable deviations up to \SI{95}{\percent}, which may have resulted in handling-induced defects (see Section~\ref{sec:sensorBackside}) by changing to backside-biasing. Currents of cells in breakdown are not considered in this analysis.

\begin{figure}[h!]\centering
	\includegraphics[width=0.5\textwidth]{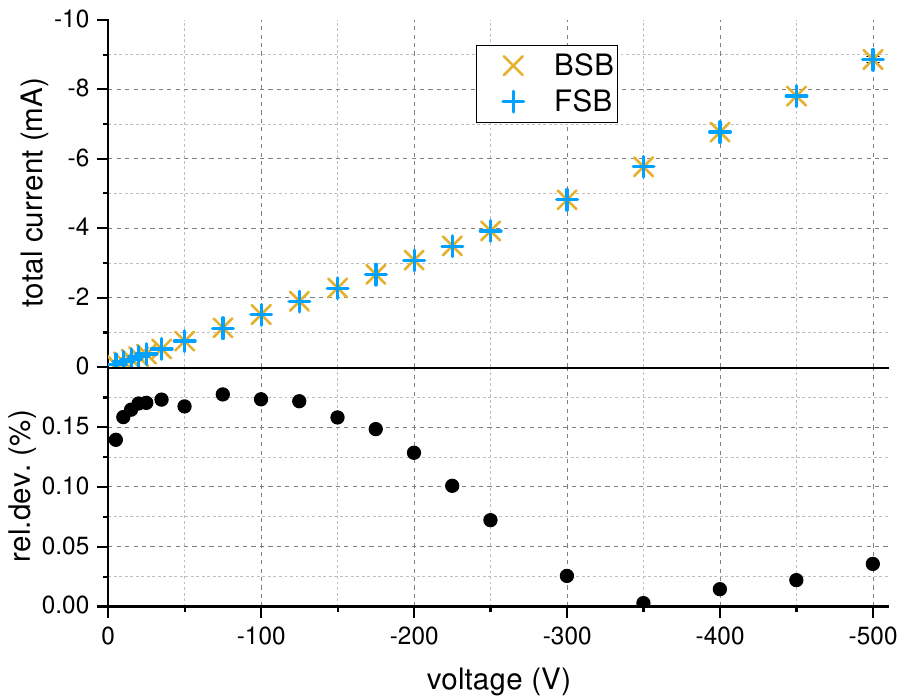}
	\caption{Total current as a function of the bias voltage, comparison between backside- and frontside biasing schemes of a \SI{200}{\micro\metre} thick sensor. Relative deviations are given by Equation~\ref{eq:ratio}.}
	\label{fig:totCurr}
\end{figure}

The measured total currents deviated by less than \SI{0.2}{\percent} between frontside and backside biasing, as shown in Figure~\ref{fig:totCurr}. Thus, we consider the frontside biasing concept as a viable option for unirradiated sensors.

\begin{figure}[H]\centering
	\includegraphics[width=0.5\textwidth]{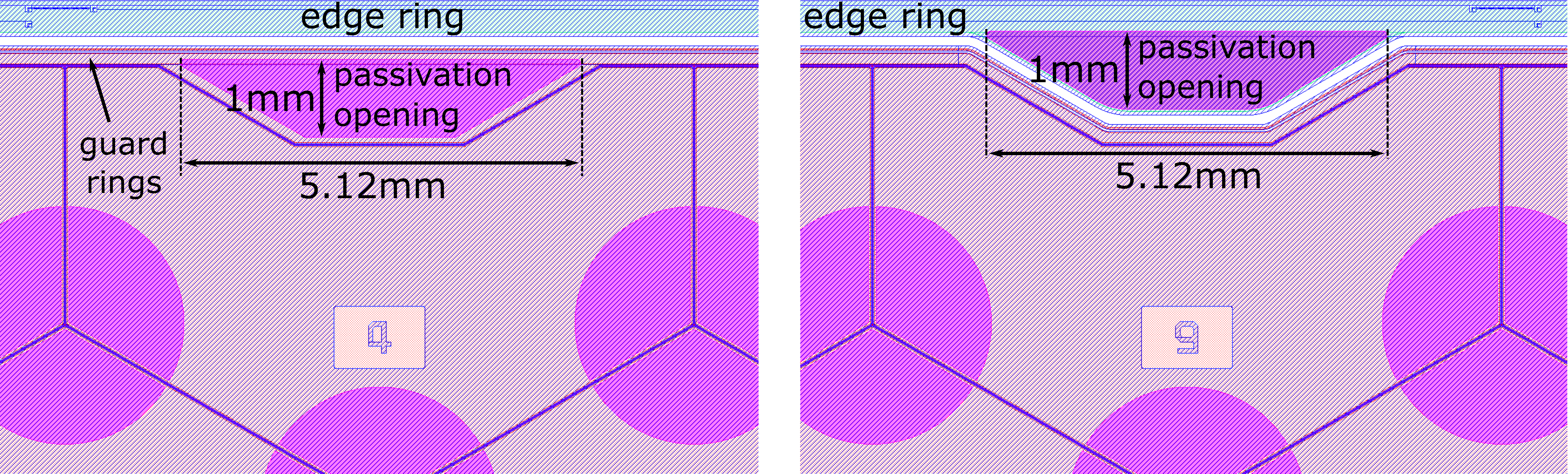}
	\caption{Extensions on the upper part of the sensor. Left: Trapezoidal guard ring extension, which allows direct contacting of the guard ring via probe-card pins. Right: Trapezoidal edge ring extension, which allows direct contacting of the edge ring via probe-card pins. The circular sectors are passivation openings for contacting the cell diodes.}
	\label{fig:extensions}
\end{figure}

In contrast to earlier sensor designs~\cite{PreeDissTS}, inward extensions of the edge ring as shown in right Figure~\ref{fig:extensions} with contact pads (passivation openings over metalization) have now been included. Since the \mbox{HGCAL} will also utilize partial sensors, we applied one edge ring extension on each long sensor edge. These structures now give the possibility to contact the edge ring via probe-cards~\cite{ARRAY}, see Chapter~\ref{subsec:probestation}. The contact area of the extensions is large enough to support two redundant pin contacts on each extension to avoid sparking in case of loss contact during the testing.

For contacting the inner guard ring, we implemented a similar concept. The main purpose of these guard ring extensions, shown in left Figure~\ref{fig:extensions}, is to ease the placement of wirebonds between the guard ring and the "Hexaboard". The Hexaboard is a printed circuit board (PCB) for power supply and readout of the silicon sensor.\ignore{ In contrast to earlier module assembly procedures, we want to guide these wirebonds through holes in the PCB to make the assembly more robust to mechanical damages.} A positive side-effect is the speedup of sensor testing because dedicated probe-card contacts using spring-loaded pins can be used instead of contacting the guard ring via separate needles. In previous designs, the guard ring was too narrow to be contacted via probe-card pins. However, \SI{0.2036}{\percent} of active sensor area gets lost due to these extensions.

	\section{Conclusions and Prospects}\label{sec:prospects}
	The aforementioned backside fragility requires adaptations of the current sensor testing and handling procedures such as the necessity of protecting the backside during testing. In this paper it was shown that frontside biasing is a practical solution to bypass the problem of biasing an insulated sensor backside.\\To qualify the radiation hardness of the new 8" process, extensive irradiation tests have to be done. Radiation changes indicators like full depletion voltage and bulk resistivity, which can be extracted from IV and CV measurements. Also, irradiation campaigns of test structures are planned which allow to investigate the radiation hardness of the  p-stop elements~\cite{MOSCATELLI2020162794}. Metal-Oxide-Semiconductor (\mbox{MOS}) test capacitors allow measuring the oxide charges via flatband voltage, which gives information about the oxide quality~\cite{Hinger20PQC}.\\It is projected to verify these results in parallel by Technology Computer-Aided Design (\mbox{TCAD}) simulations~\cite{PreeDissTS} to improve understanding the effects of the 8" process technology on radiation hardness.
	\ignore{
		Currently used neutron irradiation facilities for full sensors are the Rhode Island Nuclear Science Center\footnote{\url{http://www.rinsc.ri.gov/research/}} for 8" sensors, for 6" sensors the Jožef Stefan Institute\footnote{\url{http://www.rcp.ijs.si/ric/index-a.htm}} in Ljubljana \cite{JSIRadulovic2017}. Work-in progress is using the reactor of the Atominstitut\footnote{\url{https://ati.tuwien.ac.at/reactor/EN/}} in Vienna solely for test structures. Of course, test structures can (and are) irradiated also in the other facilities. Irradiation of test structures with electrons and protons is done at Fermilab\footnote{\url{https://fnal.gov/pub/science/particle-accelerators/accelerator-complex.html}}, while for X-ray irradiation, facilities at \mbox{CERN} and the University of Padova\cite{MOSCATELLI2020162794} are used.
	}
\section*{Acknowledgements}
	This project has received funding from the call ''Forschungspartnerschaften'' of the Austrian Research Promotion Agency (\mbox{FFG}), Austria under the grant no. 868296.
	\bibliography{hstd19Paulitsch}
\end{document}